Current-Driven Microwave Dynamics in Magnetic Point Contacts as a Function of Applied Field Angle

W. H. Rippard, M. R. Pufall, S. Kaka, T. J. Silva, and S. E. Russek

*National Institute of Standards and Technology, Boulder, CO 80305*

Abstract

We have measured microwave frequency, current-driven magnetization dynamics in point contacts made to $Co_{90}Fe_{10}/Cu/Ni_{80}Fe_{20}$ spin valves as a function of applied field strength and angle relative to the film plane. As the field direction is varied from parallel to nearly perpendicular, the device power output increases by roughly two orders of magnitude while the frequencies of the excitations decrease. For intermediate angles the excited frequency does not monotonically vary with applied current and also exhibits abrupt, current-dependent jumps. For certain ranges of current, and applied field strength and direction, the excitation linewidths decrease to a few megahertz, leading to quality factors over 18,000.

Current driven excitations in magnetic nanostructures are expected to be increasingly important as the size of magnetic-based devices continues to shrink.[1,2] At device dimensions below a few hundred nanometers, the interactions between a spin-polarized current and a thin ferromagnetic film can dominate over the effects of an externally applied magnetic field.[3] Such excitations may have negative consequences for the stability of future generation hard-disk read heads, but may also lead to new methods for current controlled switching in nanomagnetic devices such as magnetic random access memory elements. Recent experiments have also demonstrated the existence of spin transfer induced coherent high frequency microwave excitations.[4,5] This new class of microwave oscillator may have potential uses as nanometer scale high frequency sources compatible with conventional semiconductor processing.[6]

We have previously investigated these current induced excitations in magnetic point contacts as a function of applied magnetic field $H$ and current $I$, for both in-plane and out-of-plane fields.[4] Here we report on the high frequency excitations as the direction of the applied field is varied between these two extremes in order to more fully explore the range of precessional dynamics excited by the spin-torque effect. We demonstrate substantially narrower linewidths and increased output power than previously shown, both being relevant for potential technological applications. Specifically, for certain applied field geometries and currents we observe linewidths below 2 MHz and device output voltage exceeding 10 % of the maximum obtainable through the giant-magnetoresistance (GMR) effect. On average, the dependencies of the excitation frequencies on field strength and angle are similar to those in ferromagnetic resonance (FMR) measurements, as expected from theory.[1] However, their dependence on current is typically more complicated than predicted by single domain, constant damping simulations based on Ref. [1].

The studies here were carried out on a lithographically defined point contact ($\approx$ 40 nm in diameter) made to the top of a continuous 8 μm x 12 μm spin valve mesa. The spin valve structure consisted of $SiO_2$/Ta (2.5 nm)/Cu (50 nm)/$Co_{90}Fe_{10}$ (20 nm)/Cu (5nm)/$Ni_{80}Fe_{20}$ (5 nm)/Cu (1.5 nm)/ Au (2.5 nm). The $Co_{90}Fe_{10}$ is considered the "fixed" layer $M_{fixed}$ in terms of the spin-torque effect due to its larger thickness (volume) and larger saturation magnetization relative to the "free" $Ni_{80}Fe_{20}$ layer $M_{free}$. All data presented were measured on a single device having a dc resistance of 15 Ω and a GMR value of 150 mΩ, although qualitatively similar results to those presented have been observed in other devices. The device is current-biased so that changes in the relative orientations of the magnetizations of the two layers appear as voltage changes across the device due to the GMR effect. The device is contacted with microwave probes, and a bias-tee is used to separate the injected dc current and the high frequency device response. The output is amplified and measured using a 50 GHz spectrum analyzer. The gain from the amplifier has been divided out of the presented data. The center frequencies $f$ of the excitations are determined from Lorentzian fits to the measured spectra. All measurements were performed at room temperature.

In Fig. 1(a) we show the device oscillation frequencies as a function of $I$ for several field angles $\theta_H$, given relative to the film plane, for a constant field $\mu_0 H$= 0.8 T. For in-plane fields, the frequency output linearly red-shifts with current ($f$ decreases with increasing $I$), as is generally observed in these devices for in-plane fields, for all magnitudes of $H$.[4] As the angle of the applied field is increased, the excitations typically appear over a wider range of currents and the dependence of $f$ on $I$ becomes more complicated. For the data at $\theta_H = 35°$ a linear red-shift is found for low currents. However, at $I = 5$ mA the slope of the curve $df/dI$ changes sharply and, although the current-induced red-shift persists, $f$ shows significant deviations from a linear dependence on $I$. As the angle is increased, this initial sharp change in slope becomes an abrupt jump in the excitation frequency, as shown by the data for $\theta_H = 45°$, defining two distinct frequency branches in the $f$ vs. $I$ curve. For $\theta_H = 55°$ the precession frequency initially decreases but then increases (blue-shifts) with current for $I > 5$mA. At this angle, as $I$ reaches 6.625 mA the excited mode becomes poorly defined (the excitation linewidth is several gigahertz and the amplitude strongly decreases), and we were not able to uniquely determine $f$ for 6.75 mA $< I <$ 7.25 mA. However, as $I$ increases further the mode again becomes well-defined and blue-shifts with current.

As the field angle is increased, abrupt jumps in the frequency of the oscillations with increasing current are again seen. For instance, for $\theta_H = 65°$ the oscillation frequency red-shifts for currents below 6 mA, whereas for higher currents the oscillations abruptly shift to a higher frequency and show a blue-shift with increasing current. At higher angles similar multiple jumps in $f$ with $I$ are still seen but with the frequency now showing an *overall* blue-shift on each of the individual branches of the curves (see $\theta_H = 75°$). However, the frequency does not typically vary linearly, or even monotonically, with $I$ over the entire range of the individual branches of these curves. For instance, for the middle branch of the $\theta_H = 75°$ data, $f$ shows a blue-shift at low current but a red-shift for $I > 6.75$ mA. The same qualitative features described above occur over the range of fields studied (0.5 T to 1.1 T) although the particular current and angle at which two frequency branches are delineated, as well as the detailed dependence of $f$ on $I$ over a particular branch, varies with $H$. For a given angle $f$ can, on average, be tuned over a range of $\approx$ 2 GHz over the currents studied.

As shown in Fig. 1 for $\theta_H = 65°$ and $I \approx 5.75$ mA, the frequency output of the device at fixed current and field can be multivalued. This is not hysteretic behavior with $f$ depending on the direction of current sweep, but rather multiple non-harmonically related peaks are observed in the spectral output of the device at this particular current and field. Individual time-sequenced spectra often show the powers in the individual peaks change significantly from scan to scan with the power associated with one of the frequencies increasing or decreasing at the expense of the other. We attribute this to the device hopping between distinct precessional trajectories with different oscillation frequencies. Often each individual peak in a multipeak spectrum has a linewidth < 50 MHz. In some cases this hopping behavior is not explicitly observed, likely due to our $\approx$ 100 ms spectral acquisition time that limits direct detection of this switching behavior to situations in which one of the precessional states has a dwell time of that order or longer.

We performed single-domain simulations of current induced dynamics based on a Landau-Lifshitz-Gilbert equation modified to include the effects of spin torque.[1] The effects of the spin torque on both of the magnetic layers in the spin-valve as well as those of finite temperature are included. As we have noted previously [4] these simulations only approximate the present experimental geometry in which a small electrical contact is made to an extended film. The simulations were performed over the range of angles and fields experimentally investigated and show some of the qualitative features of the observed behavior discussed above. Generally, the simulations indicate that the precession of $M_{free}$ initially occurs about its equilibrium ($I = 0$ mA) direction. As $I$ increases, the precessional cone angle increases while the precessional frequency decreases.[7] As $I$ increases further, the precession acquires a time-averaged component perpendicular to its equilibrium direction in the plane defined by **H** and the equilibrium direction of $M_{free}$. In this regime, the precession frequency blue-shifts with current. As this transition from red-shift to blue-shift occurs, there is a sharp increase in the precession frequency and a large broadening of the linewidth, similar to the features seen in the $\theta_H = 55°$ data. In the simulations, at high currents $M_{fixed}$ also begins to precess and a second jump in the precession frequency of the free layer occurs. The effect of the fixed layer precession is to increase the precession frequency of $M_{free}$. While the measured jumps in $f$ can possibly be equated with such behavior, many features of the measured device response are not found in the simulations. For instance, in the simulations no geometry and current yields multiple excitation frequencies. Experimentally, particular angles and fields can yield as many as four distinct branches in $f$ vs. $I$ curves. However, only a maximum of three have so far been found in the simulations, a red-shifting branch and two blue-shifting branches (corresponding to static and precessional motion of $M_{fixed}$). Additionally, we have measured frequency jumps to *lower* precessional frequencies with increasing $I$ (see $\theta_H = 45°$ data), which are not seen in the simulations for any geometry studied. These discrepancies possibly indicate limits to the applicability of single-domain simulations having constant damping to properly model point-contact experiments.

In Fig. 1b we show the linewidths $\Delta f$ corresponding to the data in Fig. 1a, where $\Delta f$ is the full-width-at-half-maximum (FWHM) of the spectrum in units of power. For clarity only three different field angles are presented but are representative of the entire data set. As shown in the figure, the linewidths have complicated dependencies on both current and field angle. Overall the linewidths can vary by nearly two orders of magnitude even for a single applied field direction and strength. The linewidths are typically not simple monotonic functions of $I$ and often show changes correlated with features in the $f$ vs. $I$ curves. For instance, the linewidths for $\theta_H = 45°$ are < 10 MHz at the lowest currents, then increase to > 50 MHz at $I = 4.75$ mA before dropping again to about 2 MHz and then gradually increasing with increasing $I$. In this case the abrupt change in the linewidth correlates with a jump in $f$ with current. Between $I = 4.75$ mA and 5.0 mA, the frequency changes from 24.7 GHz to 24.3 GHz. The abrupt changes in linewidth for the data at $\theta_H = 75°$ also correspond to abrupt changes in the oscillation frequency. However, in this case the linewidth increases at $I = 4.625$ mA but decreases at $I = 7.5$ mA. In many cases, the excitation linewidth undergoes a significant increase prior to these frequency jumps. However, this is not always the case, as seen for the $\theta_H = 75°$, $I = 4.5$ mA data. Occasionally an abrupt frequency shift is not accompanied by any significant change in linewidth, as shown by the data for $\theta_H = 65°$ and $I = 6$ mA. The average linewidth for all of the data in Fig. 1a is 17.5 MHz.

Figure 2 shows the precession frequency as a function of both the angle and strength of the applied field. The data points represent the mean oscillation frequency over a current range of 3 mA to 8 mA for a given $\theta_H$ and $H$. For small angles and fixed $H$, the frequencies change only moderately with angle. As the field angle is increased, $df/d\theta$ is negative with monotonically increasing magnitude. The variation of frequency with $H$ also changes with the angle of applied field. For instance, for $\theta_H = 10°$ we find

d$f$/d($\mu_0 H$) = 27 ± 0.5 GHz/T, in good agreement with our previous measurements for in-plane fields.[4] As the field angle is increased d$f$/d($\mu_0 H$) decreases, reaching a value of 10 ± 0.4 GHz/T at $\theta_H$ = 85°. For comparison, we calculated the FMR frequencies [8] of our device for a number of different angles and applied fields, assuming $\mu_0 M_{CoFe}$ = 1.8 T, $\mu_0 M_{NiFe}$ = 0.9 T, and a Landé factor $g$ = 2 (see line on Fig. 2), which are the same frequencies observed in the simulations for low currents. We find qualitative agreement between the measured and calculated trends in both d$f$/d$H$ and d$f$/d$\theta_H$. For all fields, the values of the calculated frequencies at small field angles are about 20 % higher than those measured, whereas at large angles they are about 10 % lower than the measured signals. This behavior is consistent with the precession angle of the oscillations being significantly larger than the small angle approximation used for FMR calculations. Unlike in our previous measurements of a different device [4], for $\theta_H$ = 90° the frequency and power output of this device becomes highly hysteretic in both $I$ and $H$, making a detailed discussion problematic. This onset of hystersis may be due to a lack of in-plane anisotropy in the "free" layer, or to a particular physical and/or magnetic configuration directly under the contact area. All other qualitative features reported here have been confirmed to be present in other devices although the particular currents, frequencies, and associated linewidths vary from device to device.

For a particular field strength and angle the device output power is typically a strong function of $I$, Fig. 3a. The power output generally does not scale as $I^2$, but depends on the particular frequency branch of the excitation, consistent with the trajectories of $M_{free}$ changing with current. Shown in Fig. 3b is the *maximum* integrated power output of the device vs. applied field angle. In general, the current yielding the maximum power output varies with $H$ and $\theta_H$. For this device this current varies between $I = 5$ mA and 7 mA. Hence, normalizing the data by $I^2$ does not significantly affect the trend in the plot. The maximum power output of the device is a strong function of both $H$ and $\theta_H$, varying by roughly two orders of magnitude, from about 1 pW to 0.1 nW, as the field is changed from in-plane to out-of-plane. For simple circular precession, the GMR signal should follow $\Delta R = \Delta R_{max} \sin(\gamma)\sin(\beta)$ where $\Delta R_{max}$ is the maximum MR signal, $\gamma$ is the angle between the time averaged values of $M_{fixed}$ and $M_{free}$ (generally different from $\theta_H$, Fig. 3c), and $\beta$ is the precession angle (inset Fig. 2). Hence, for a constant precession angle, the device power output should monotonically increase with $H$ at a given $\theta_H$ and roughly scale as $\sin(\gamma)$ at fixed current, over the range of fields studied here. While the measured power output does increase with $\theta_H$, its dependence on angle at fixed $H$ does not follow such a simple relation (solid line in Fig. 3b). Furthermore, the power output generally does not simply increase with $H$ at a given applied field angle, indicating that the excited trajectories are more complicated functions of $H$ and $I$.

The excitation linewidths also strongly vary with current as well as the applied field direction and strength, and are quite narrow for particular geometries. For example, Fig. 4a shows the excitation spectrum for the device at $\theta_H = 30°$. A Lorentzian fit to the data yields $f = 34.38$ GHz and $\Delta f = 1.89$ MHz, leading to a quality factor for the oscillation $Q = f/\Delta f = 18,200$. Narrow linewidths are not exclusive to a particular applied field direction and strength (see Fig. 1b) but typically occur over a range of currents and field strengths for a particular direction of the applied field. Moreover, the existence of a narrow linewidth excitation does not presume small angle (low power output) precession. For instance, Fig. 4b shows the spectral output of device with $\theta_H = 85°$. The integrated output power due to the excitation is 86 pW, corresponding to a peak voltage of $V = 93$ µV while the linewidth of the excitation is only $\Delta f = 3.2$ MHz. For comparison, the maximum possible MR derived voltage output of the device for this current is 900 µV.

In summary, we have measured the frequency and power dependencies of current induced excitations in point contacts as a function of applied field angle and strength. For intermediate field angles, the precessional frequency and device output power show complicated dependencies on current. Abrupt jumps in the excitation frequency are found as well as the existence of multiple stable precessional states at particular fields and currents. The power output increases by roughly two orders of magnitude as the field is varied from in-plane to out-of-plane, as expected from simple geometrical arguments. However, the dependence of output power is not monotonic in either field angle or current, indicating that the excited trajectories have complicated dependencies on current and field. In certain geometries, the linewidth of the excitation is below 2 MHz leading to oscillations with $Q > 18,000$. Furthermore, such narrow linewidths are not limited to small angle (low power) precessional modes. Many of the qualitative behaviors found in our measurements are not found in single domain simulations.

Work supported by the DARPA SPinS and the NIST Nanomagnetodynamics programs. We thank M. D. Stiles and F. B. Mancoff for helpful comments.

Figure Captions

Fig. 1 (a) Frequency vs. current for several different field angles for $\mu_o H = 0.8$ T. The FWHM of the spectra are smaller than the data points. (b) Linewidths associated with data shown in part (a). Both increasing and decreasing $I$ scans are shown but are not visible on this scale range.

Fig. 2 Average $f$ vs. $\theta_H$ for several values of $H$. Error bars represent the maximum and minimum $f$ excited over the range $I = 3$ mA to 8 mA. Error bars not shown at large angles for clarity but are on average ± 0.6 GHz. Solid line shows calculated FMR frequencies for $\mu_0 H = 0.8$ T. Inset shows angles discussed in the text.

Fig. 3 (a) Integrated output power (area under spectral peak) vs. $I$. For currents having multiple frequencies the powers in both peaks are included. Data correspond to those of the same symbol in Fig. 1(a). (b) Maximum integrated power output vs. field angle for several different fields. The line slows the calculated functional form of the power for constant circular precessional angle and $\mu_o H = 0.8$ T. (c) Calculated values of $\gamma$ for several fields.

Fig. 4 (a) Spectral output showing a narrow linewidth and high $Q$ value. (b) Spectral output in a different field geometry showing a high output power state. For both figures $I = 6$ mA. Solid lines are Lorentzian fits.

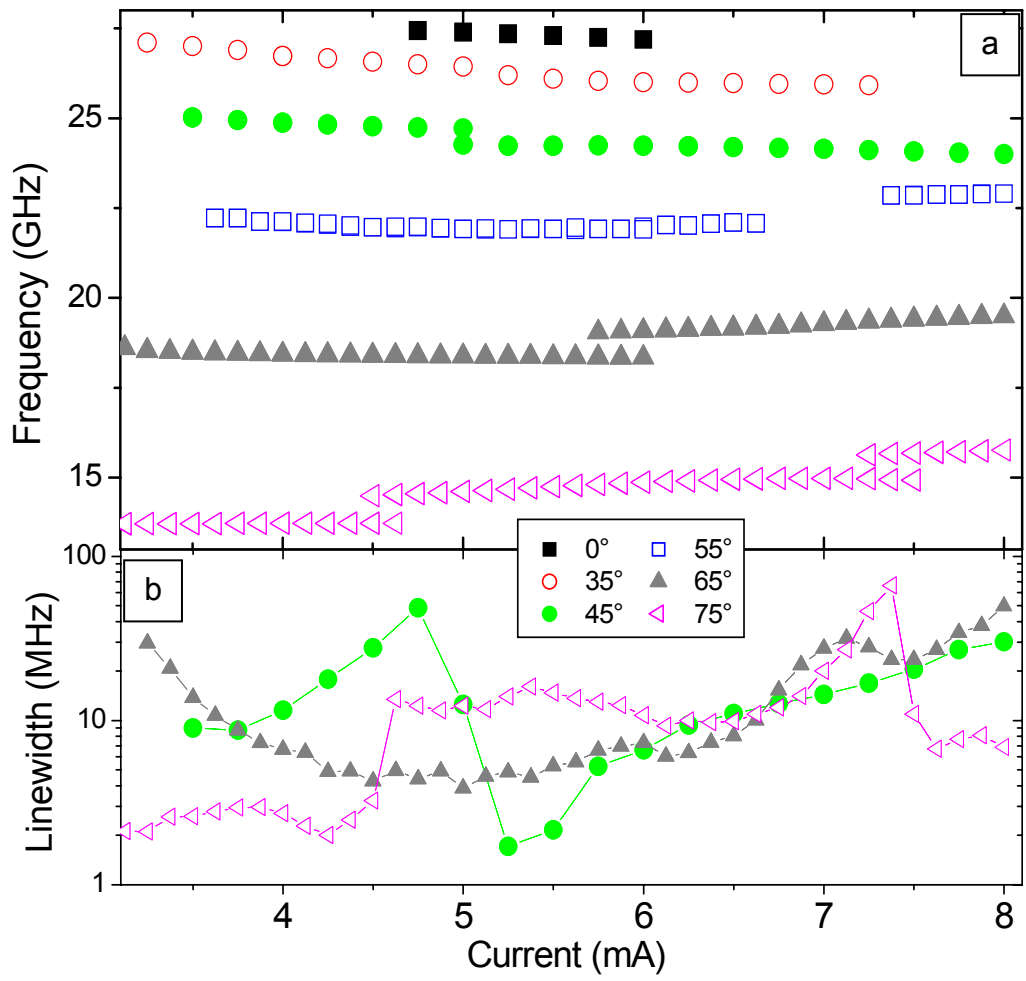

Fig. 1

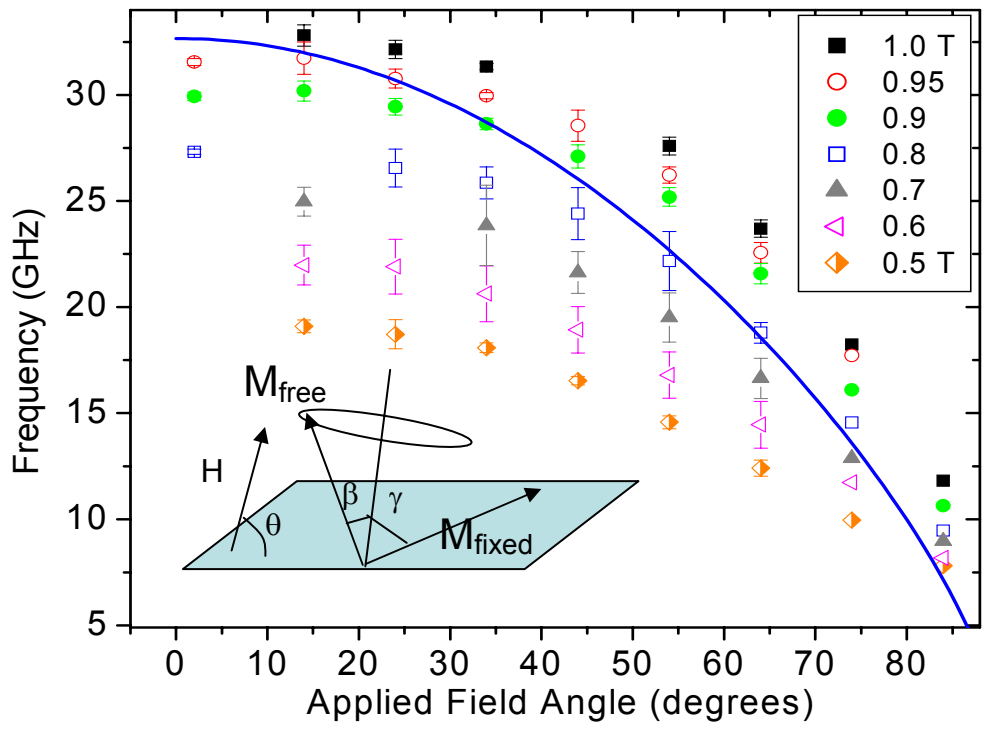

Fig. 2

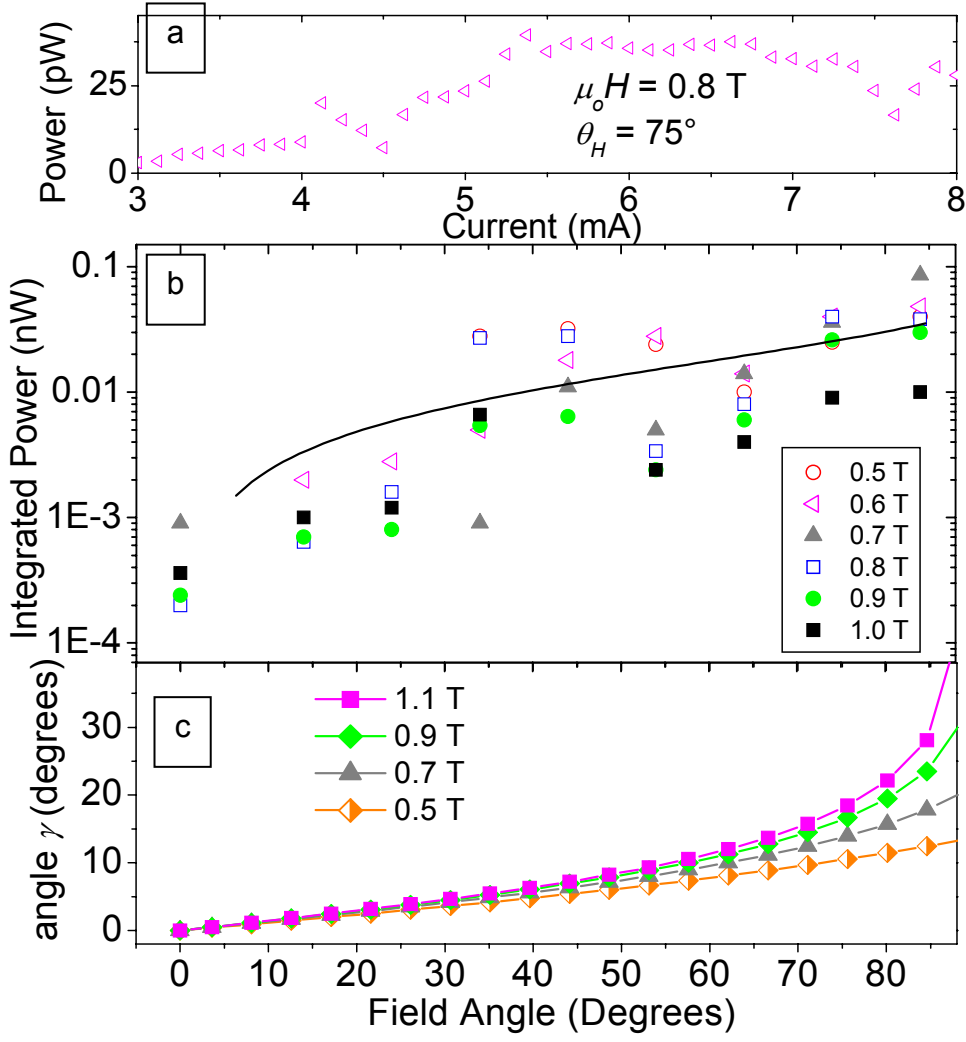

Fig 3

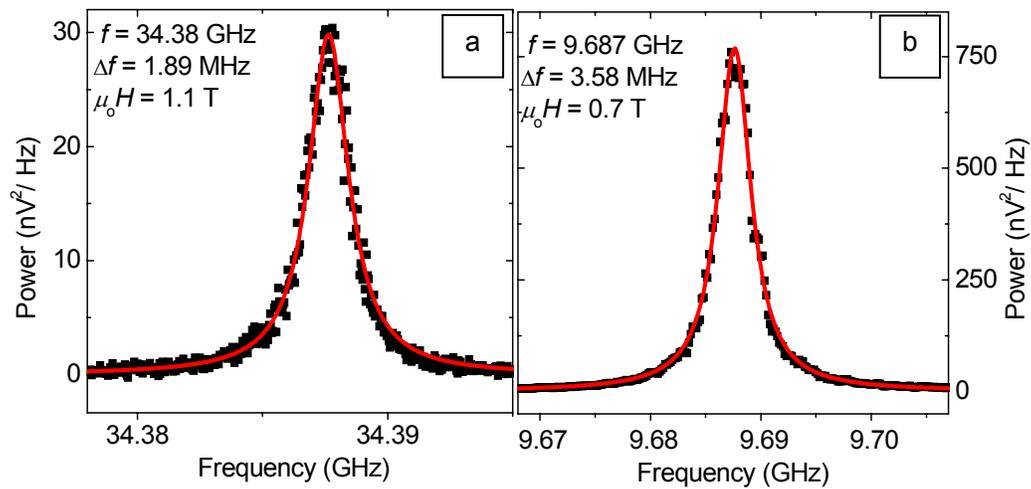

Fig. 4